\documentclass[11pt]{article}
\usepackage{moriond,epsfig}
\usepackage{cite}
\usepackage{mathtools}
\usepackage{amssymb}
\usepackage{wasysym}
\usepackage{paralist}
\usepackage[usenames,dvipsnames]{color}
\usepackage{subfigure}

\usepackage{hyperref}
\hypersetup{
    bookmarks=false,      % show bookmarks bar?
    pdfstartview={FitH},  % fits the width of the page to the window
    colorlinks=true,      % false: boxed links; true: colored links
    linkcolor=black,      % color of internal links
    citecolor=blue,       % color of links to bibliography
    filecolor=blue,       % color of file links
    urlcolor=blue         % color of external links
}

\def\be{\begin{equation}}
\def\ee{\end{equation}}
\def\bea{\begin{eqnarray}}
\def\eea{\end{eqnarray}}

\newcommand{\SU}[1]{\ensuremath{\text{SU}(#1)}}
\newcommand{\sgn}[1]{\ensuremath{\text{sgn\,}\mu}}
\newcommand{\gut}{{\textsc{gut}}}
\newcommand{\mgut}{M_{\gut}}

\begin{document}
{\hfill LPSC-11111}
\vspace*{4cm}
\title{A FOURTH CHIRAL GENERATION AND SUSY BREAKING}

\author{ A.~WINGERTER }

\address{Laboratoire de Physique Subatomique et de Cosmologie\\
UJF Grenoble 1, CNRS/IN2P3, INPG\\
53 Avenue des Martyrs, F-38026 Grenoble, France}

\maketitle\abstracts{
We revisit four generations within the context of supersymmetry. We
compute the perturbativity limits for the fourth generation Yukawa
couplings and show that if the masses of the fourth generation lie
within reasonable limits of their present experimental lower bounds,
it is possible to have perturbativity only up to scales around 1000
TeV, i.e.~the current experimental bounds and perturbative unification
are mutually exclusive. Such low scales are ideally suited to
incorporate gauge mediated supersymmetry breaking, where the mediation
scale can be as low as 10-20 TeV. The minimal messenger model,
however, is highly constrained. Lack of electroweak symmetry breaking
rules out a large part of the parameter space, and in the remaining
part, the fourth generation stau is tachyonic. Contribution to the
proceedings of \textit{Les Rencontres de Moriond EW 2011} based on
ref.~\cite{Godbole:2009sy}.
}

\section{Is There Room For a Fourth Generation?}

It was long believed to be common lore that an extra chiral generation of fermions was excluded by electroweak precision measurements \cite{Amsler:2008zzb}. Recently, however, the interest in a fourth sequential generation of fermions was revived after realizing that the constraints are by far not as stringent as they were thought to be. We start out by reviewing what we know about the fourth generation couplings to Standard Model quarks and leptons, and then revisit the various constraints from experiment and theory.

\subsection{Determination of $\left|V_{tb}\right|$}

Until recently the CKM matrix element $\left|V_{tb}\right|$ was determined indirectly from the ratio \cite{Abazov:2006bh}
\begin{equation}
R = \frac{\mathcal{B}(t\rightarrow Wb)}{\mathcal{B}(t\rightarrow Wq)} = \frac{|V_{tb}|^2}{|V_{td}|^2+|V_{ts}|^2+|V_{tb}|^2} = 1.03^{+0.19}_{-0.17}
\end{equation}
by assuming that $|V_{td}|^2+|V_{ts}|^2+|V_{tb}|^2=1$, or in other words, that the $3\times3$ CKM matrix is unitary. A value $R\neq1$ would either imply non-standard-model-like interactions of the top quark or the existence of extra quarks \cite{Amsler:2008zzb}. From $R=1$, however, we cannot infer that a fourth generation is ruled out \cite{Alwall:2006bx}. The first direct measurement from observing single-top production \cite{Abazov:2009ii,Aaltonen:2009jj} gave $\left|V_{tb}\right| > 0.78$ which is consistent with $\left|V_{tb}\right| = 1$, but leaves enough room e.g.~for a fourth generation replica of the top quark, $t'$, to couple to the $b$ quark.

\clearpage
\subsection{Do Electroweak Precision Observables Really Forbid a Fourth Generation?}
\label{sec:ewobserv}

\begin{figure}[h!]
\begin{center}
\includegraphics[height=3in]{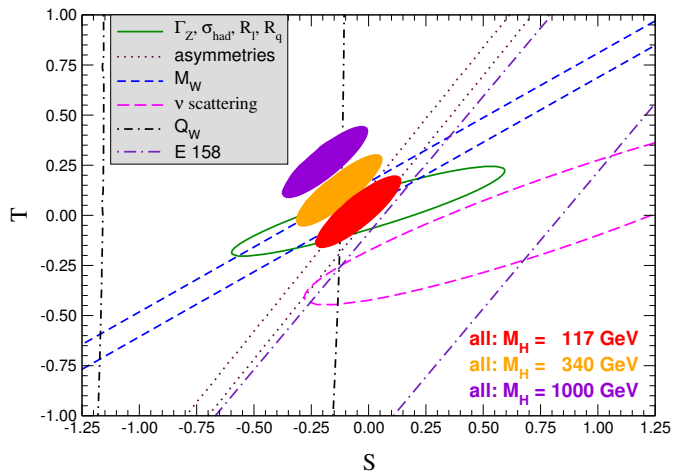}
\end{center}\vspace{-3ex}
\caption{The 90\% C.L.~contours in the $S$ and $T$ plane for different Higgs masses (by J.~Erler and P.~Langacker, see ref.~\cite{Amsler:2008zzb}). If the new physics contribution is such that $\Delta S\simeq\Delta T > 0$ is not too large, we have a fair chance of staying inside the confidence level ellipses, but we have to assume a larger Higgs mass.\hfill\quad}
\label{fig:EWPrecision}
\end{figure}

In Fig.~(\ref{fig:EWPrecision}) we show the constraints on new physics from the electroweak precision data. The contribution of a fourth chiral generation to the $S$ and $T$ parameters can be arranged to stay inside the \textit{solid} confidence level ellipses by carefully adjusting the fermion masses so that $\Delta S$ and $\Delta T$ are positive, roughly equal and not too large, and assuming a \textit{larger Higgs mass} \cite{Kribs:2007nz}. The constraints coming from electroweak precision data can thus be circumvented.

\vspace{-5ex}

\subsection{What About the Other Constraints?}

Constraints from FCNCs and $b\rightarrow s\gamma$ \cite{Bobrowski:2009ng} are typically weaker than those from electroweak precision tests \cite{Chanowitz:2009mz}. From the invisible decay width of the $Z$ boson we know that there are exactly three neutrinos that couple to the $Z$ and are lighter than $M_Z/2$, so for the fourth generation neutrino we have to assume $m_{\nu4}\geq M_Z/2$. The limit from cosmology $\sum m_{\nu} \lesssim 1$ eV does not hold for heavy neutrinos \cite{Dolgov:2002wy}.

\subsection{Our Theoretical Prejudice}

From a theoretical point of view it certainly looks odd that the fourth generation neutrino should be so much heavier than those from the first three generations. In absence of a complete and convincing neutrino mass generation mechanism, however, we cannot simply dismiss the idea on theoretical grounds. 

Asymptotic freedom of QCD restricts the number of generations to be $\leq8$, but there is no really good reason why it should be exactly three. There are e.g.~models in string theory that relate the number of generations to topological invariants of the compactification manifold, but these models usually fail to describe the details of the low-energy theory in a predictive way. 

From direct searches at colliders we know that the fourth generation fermion masses are much larger as compared to their third generation counterparts, and one may be worried about the Yukawa couplings becoming non-perturbative. Loss of perturbativity simply means that our calculations become more cumbersome and has no bearing on the validity of the theory. In that case, the fourth generation fermions may form a condensate, see e.g.~\cite{soni:20110317}.

% \begin{table}[t]
% \caption{Experimental Data bearing on $\Gamma(K \ra \pi \pi \gamma)$
% for the $\ko_S, \ko_L$ and $K^-$ mesons.\label{tab:exp}}
% \vspace{0.4cm}
% \begin{center}
% \begin{tabular}{|c|c|c|l|}
% \hline
% & & & \\
% &
% $\Gamma(\pi^- \pi^0)\; s^{-1}$ &
% $\Gamma(\pi^- \pi^0 \gamma)\; s^{-1}$ &
% \\ \hline
% \mco{2}{|c|}{Process for Decay} & & \\
% \cline{1-2}
% $K^-$ &
% $1.711 \times 10^7$ &
% \begin{minipage}{1in}
% $2.22 \times 10^4$ \\ (DE $ 1.46 \times 10^3)$
% \end{minipage} &
% \begin{minipage}{1.5in}
% No (IB)-E1 interference seen but data shows excess events relative to IB over
% $E^{\ast}_{\gamma} = 80$ to $100MeV$
% \end{minipage} \\
% & & &  \\ \hline
% \end{tabular}
% \end{center}
% \end{table}

\section{Why Is A Fourth Generation Interesting?}

As already mentioned in Section \ref{sec:ewobserv}, the effect of a fourth generation with a judicious choice of masses can be such that the electroweak precision observables would prefer a larger Higgs mass (see Fig.~\ref{fig:EWPrecision}). This is very welcome, since a certain tension has built up between the best-fit value of the Higgs mass $m_H=80^{+30}_{-23}$ GeV \cite{Nakamura:2010zzi} as obtained from electroweak precision data and the direct exclusion limit of $m_H>114.4$ GeV from LEP.

If the CKM matrix is extended to a 4-by-4 matrix, there are 2 additional phases, and this extra CP violation may be large enough to make electroweak baryogenesis viable. In the Standard Model, the first order phase transition is not strong enough to preserve the generated baryon asymmetry from being washed out, but the presence of extra scalars in the MSSM may solve this problem \cite{Ham:2004xh,Fok:2008yg}. There have been speculations \cite{Hou:2008xd} that this may work even without SUSY, but this has been contradicted by ref.~\cite{Fok:2008yg}.

\section{Four Generations and Supersymmetry}

The generalization of the MSSM to four generations (MSSM4) is straightforward, since the fourth generation is an exact replica of the third one except for the larger masses. In the following, we will denote the fourth generation quarks and leptons by $t'$, $b'$, $\tau'$, and $\nu_{\tau}'$. In the graphs, we may use the alternate notation $t_4$, $b_4$, $\tau_4$, and $\nu_{\tau4}$ for the sake of better readability.

To reduce the large number of soft parameters that are introduced into the Lagrangian for supersymmetry breaking, we will look at two unifying frameworks, namely the constrained MSSM and minimal gauge mediation. Before we do that, however, we need to discuss the regime of perturbativity for the MSSM4.

\subsection{Perturbativity of the Yukawa Couplings}

\begin{figure}[h!]
\begin{center}
\subfigure[Values of $m_{t'}$, $m_{b'}$ where $h_{t'}$ becomes\newline non-perturbative for $m_{\tau'}=100.8$ GeV.]{\includegraphics[width=0.45\textwidth]{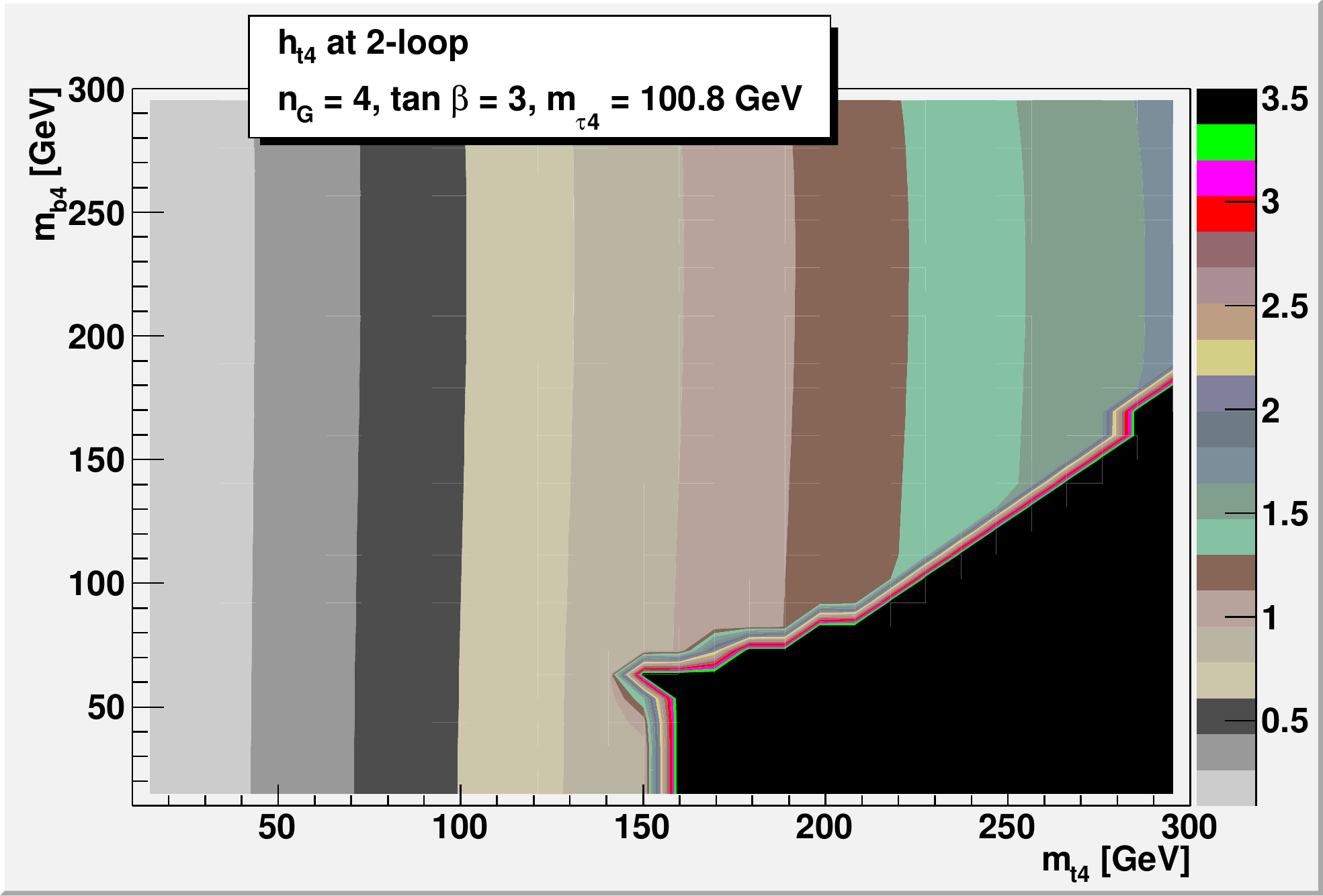}}
\subfigure[Values of $m_{\tau'}$, $m_{b'}$ where $h_{t'}$, $h_{b'}$ or $h_{\tau'}$\newline becomes non-perturbative for $m_{t'}=150$ GeV.]{\includegraphics[width=0.45\textwidth]{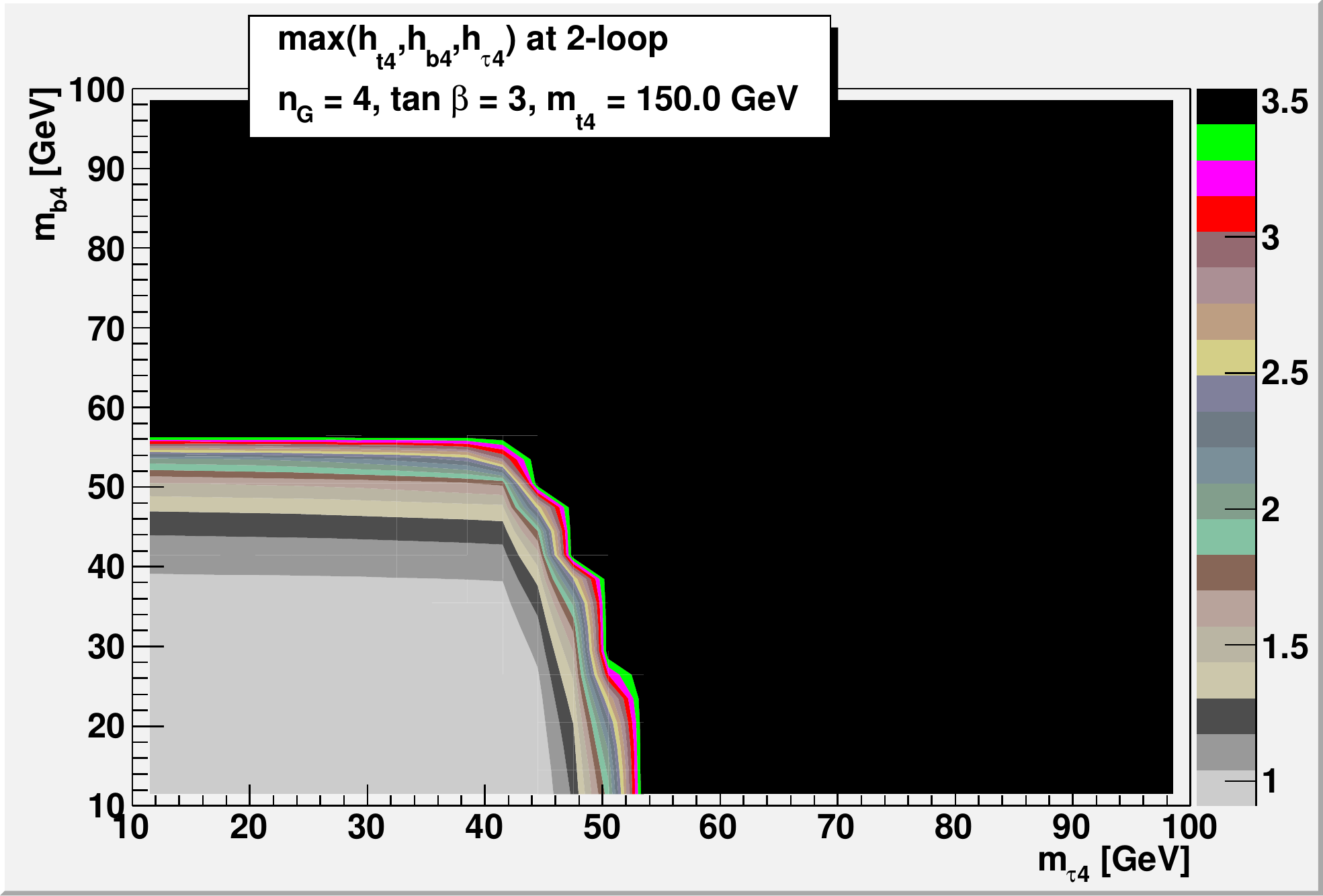}}
\caption{The black area shows the region in parameter space where the theory becomes non-perturbative.\hfill\quad}
\label{fig:pertboundmasses}
\end{center}
\end{figure}

The top Yukawa coupling is already dangerously close to the non-perturbative regime so that we have to worry about perturbation theory breaking down for not-too-large values of the fourth generation fermion masses. The left panel of Fig.~\ref{fig:pertboundmasses} shows the values of $m_{t'}$ and $m_{b'}$ for which the $t'$ Yukawa coupling becomes non-perturbative (black area) before reaching the assumed unification scale of $M_X=2.3\times10^{16}$ GeV for constant $m_{\tau'}=100.8$ GeV which is the lower experimental bound (see Eq.~(\ref{eq:explimit})). For $m_{t'}\gtrsim150$ GeV perturbativity is lost for relatively small values of $m_{b'}$.

We now fix $m_{t'}=150$ GeV and show in the right panel of Fig.~\ref{fig:pertboundmasses} the values of $m_{\tau'}$ and $m_{b'}$ for which either of the fourth generation Yukawa couplings becomes non-perturbative. The result is that $m_{\tau'}$ and $m_{b'}$ cannot be larger than roughly 50 GeV. This is in conflict with the experimental bounds \cite{Amsler:2008zzb}
\begin{equation}
m_{t'} > 256 \text{ GeV,} \quad m_{b'} > 128 \text{ GeV,} \quad m_{\tau'} > 100.8 \text{ GeV,}.
\label{eq:explimit}
\end{equation}

\begin{figure}[h!]
\begin{center}
\subfigure
{\includegraphics[width=0.45\textwidth]{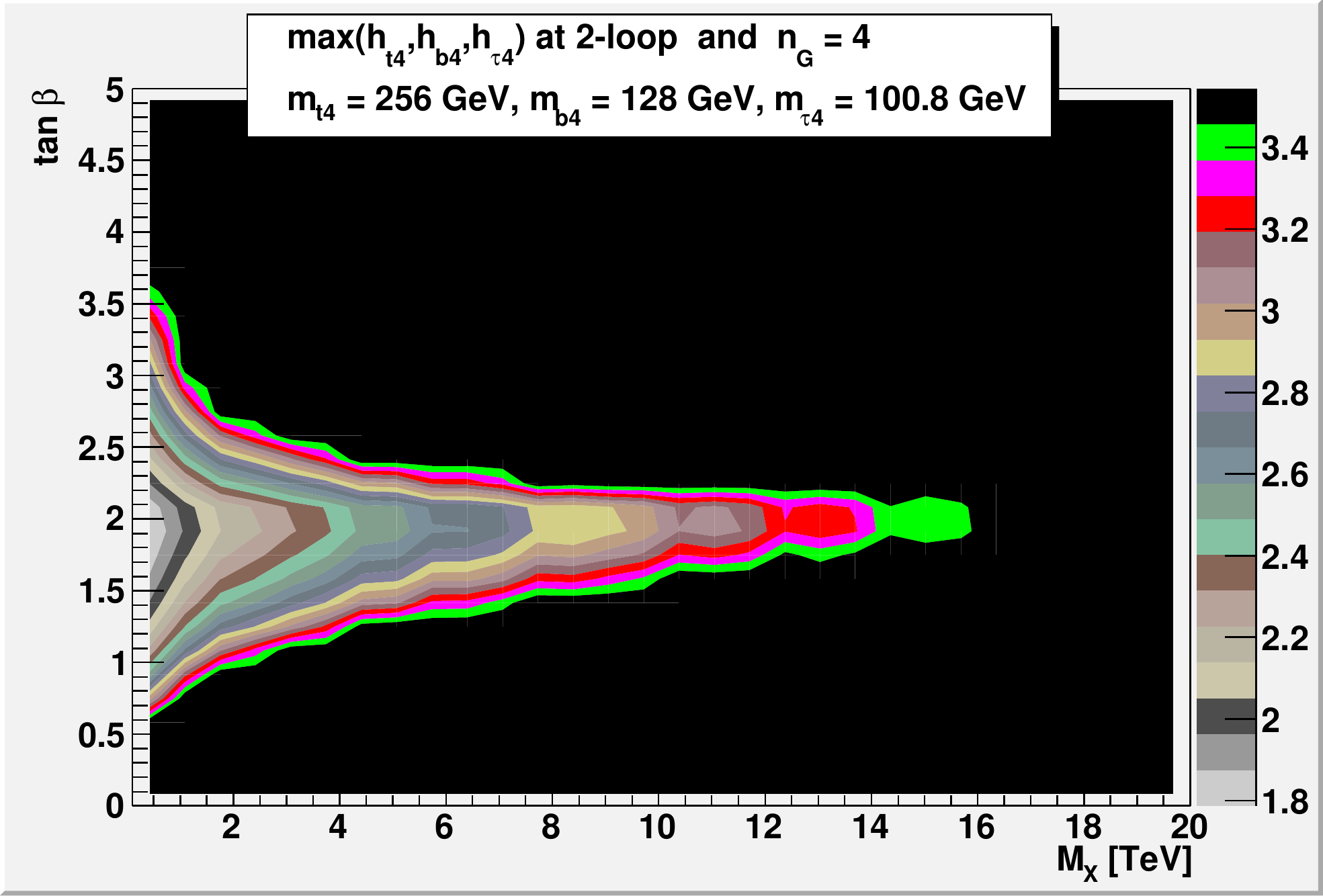}
\includegraphics[width=0.45\textwidth]{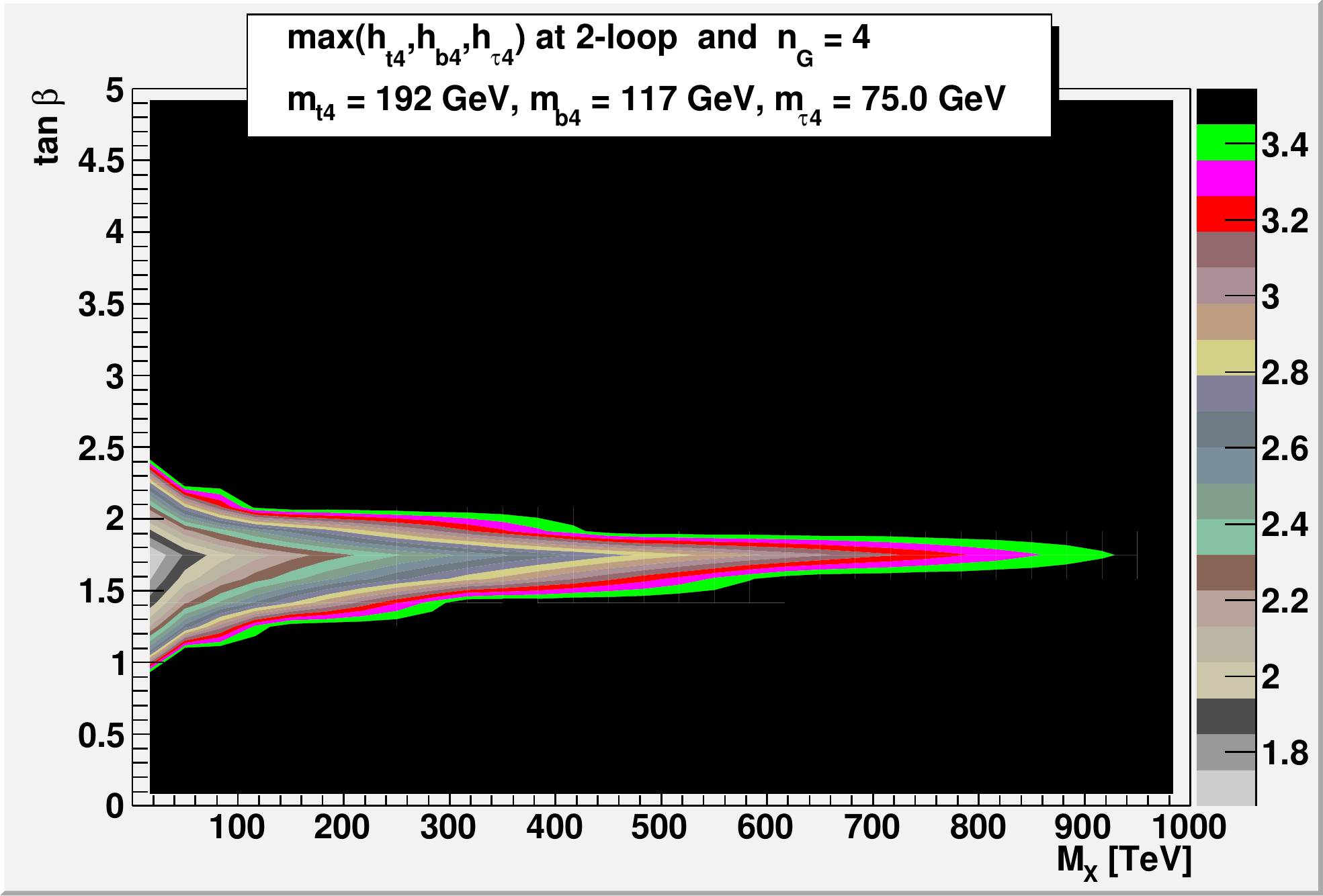}}
\caption{Perturbativity limits for the masses given in Eq.~(\ref{eq:explimit}) (left panel) and Eq.~(\ref{eq:ourlimits}) (right panel), respectively.\hfill\quad}
\label{fig:pertboundtanb}
\end{center}
\end{figure}

Reversing the logic, we fix the fourth generation fermion masses at their experimental lower bounds and ask at what scale the theory becomes non-perturbative. The left panel of Fig.~\ref{fig:pertboundtanb} shows that the highest scale that we can reach before perturbation theory breaks down is $M_X\simeq15$ TeV for $\tan\beta\simeq2$.

We can have a larger domain of perturbativity for lower fourth generation fermion masses. To that end, we recall that the limits quoted in Eq.~(\ref{eq:explimit}) are at 95\% C.L.~and can be relaxed by assuming a higher level of confidence. Furthermore, the exclusion limits denote the pole masses, whereas in our calculations, we are using the running masses. We will account for these differences by taking 25\% off the masses in Eq.~(\ref{eq:explimit}). Additionally, in order to satisfy the $T$-parameter constraint which measures the mass splitting in the \SU{2} multiplet, we impose $\left|m_{t'}-m_{b'}\right|=75$ GeV:
\begin{equation}
m_{t'} = 192 \text{ GeV,} \quad m_{b'} = 117 \text{ GeV,} \quad m_{\tau'} = 75 \text{ GeV}
\label{eq:ourlimits}
\end{equation}

Even with all these assumptions, we can maintain perturbativity only up to $M_X\simeq900$ TeV. As an aside, we remark that assuming the extra generation to be vectorlike largely avoids problems with perturbativity; we are not considering this option here, because we are interested in a \textit{sequential} extra generation that is an exact replica of the third one except for the larger masses.

\subsection{The Constrained MSSM}

The results from the previous section clearly show that the existence of a fourth generation and the idea of perturbative unification are mutually exclusive. Yet, to illustrate how things change in the presence of an extra generation, we present a \textit{toy model} where we take all fourth generation fermion masses to be equal to their third generation counterparts. 

\begin{figure}[h!]
\centering
\subfigure[Three generations.]{\includegraphics[width=0.45\textwidth]{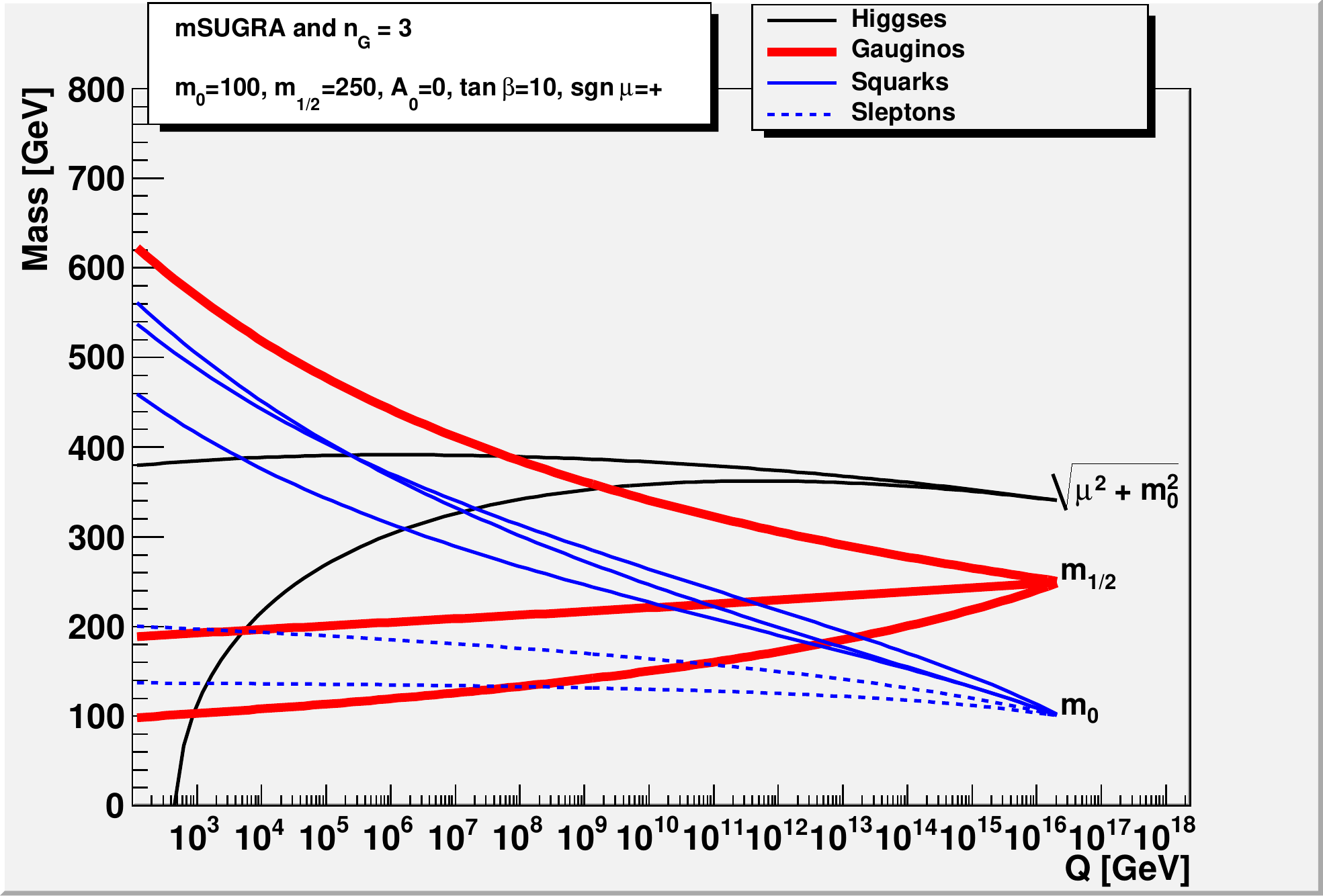}}
\subfigure[Four generations.]{\includegraphics[width=0.45\textwidth]{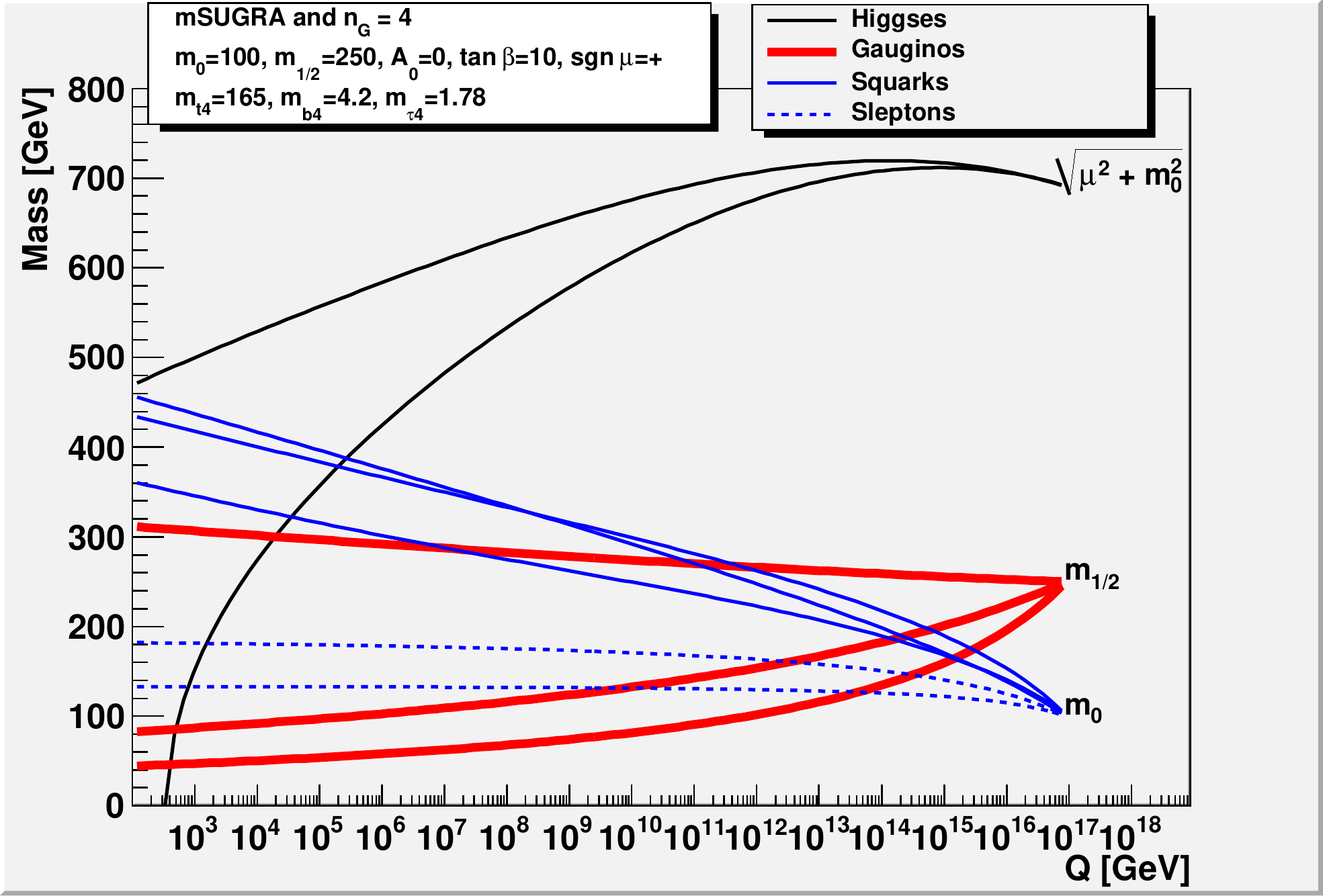}}
\caption{The running of the various soft masses in the MSSM3 (left panel) and MSSM4 (right panel). The unification scale is $\mgut{} = 2.40\times10^{16}$ GeV and $\mgut{} = 8.82\times10^{16}$ GeV, respectively.\hfill\quad}
\label{fig:4gtoyrun}
\end{figure}

In the left and right panels of Fig.~\ref{fig:4gtoyrun} we show the running of $m_0$, $m_{1/2}$ and $\sqrt{\mu^2+m_0^2}$ in the MSSM with three and four generations, respectively. We notice that in the case of four generations
\begin{inparaenum}[(i)]
\item the unification scale has increased to $\mgut{}=8.82\times10^{16}$ GeV,
\item $\left|\mu\right|$ is larger and leads to a heavier Higgs,
\item the squark and gluino masses grow slower (reading the graph from right to left).
\end{inparaenum}

These qualitative features (except the jump in $\left|\mu\right|$ in Fig.~\ref{fig:4gtoyrun}) can be easily understood by looking at the changes to the renormalization group equations in going from three to four generations. Comparing the spectra (see ref.~\cite{Godbole:2009sy}) of the MSSM3 and MSSM4 with the same mSUGRA boundary conditions, we see that the mass of the lightest CP even Higgs boson $h^0$ has indeed increased as was anticipated (the exclusion limit \cite{Nakamura:2010zzi} for a neutral MSSM Higgs boson is $m_{h^0}>92.8$ GeV). The squarks, sleptons and gluinos are lighter and the CP odd and charged Higgses are heavier.

\subsection{Minimal Gauge Mediation}

Minimal gauge mediation is especially suited for constraining the soft SUSY breaking parameters of the MSSM4, because it does not require the perturbativity of the theory to hold all the way up to the \gut{} scale. In particular, the mediation scale can be as low as 10-20 TeV.

\begin{figure}[h!]
\centering
{\includegraphics[width=0.6\textwidth]{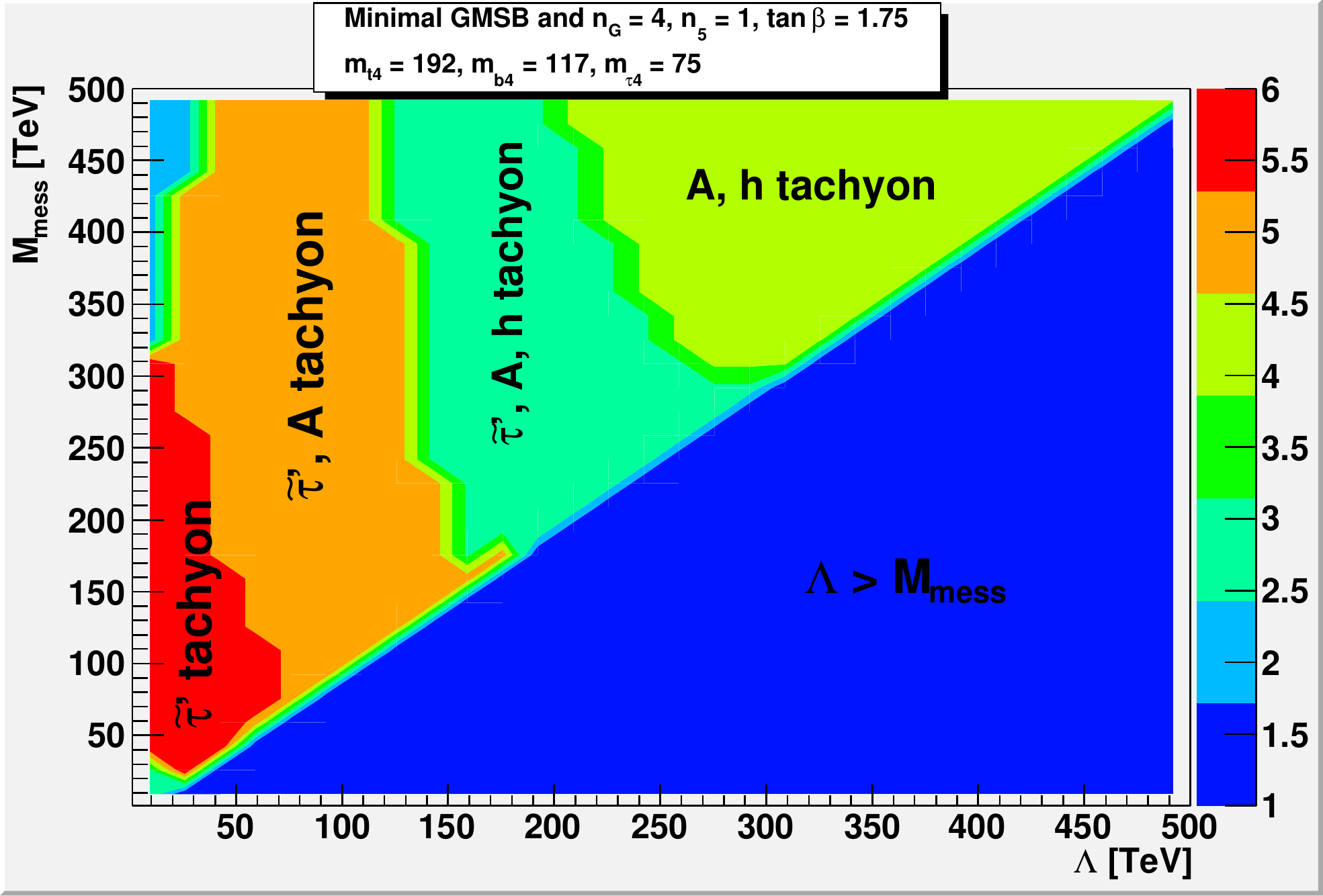}}
\caption{Regions in mGMSB parameter space $\Lambda$--$M_{\text{mess}}$ for fixed $n_5=1$, $\tan\beta=1.75$, \sgn{\mu}=+. The fourth generation masses are $m_{t'}=192$ GeV, $m_{b'}=117$ GeV, $m_{\tau'}=75$ GeV. The lower-diagonal part is ruled out as $\Lambda > M_{\text{mess}}$. In the upper-diagonal part, most of the parameter space does not have consistent radiative electroweak symmetry breaking as indicated by the tachyonic Higgses.\hfill\quad}
\label{fig:gGMSB}
\end{figure}

\clearpage

In Fig.~\ref{fig:gGMSB} we show a slice of the parameter space of minimal gauge mediation where we have fixed three of the five parameters ($n_5=1$, $\tan\beta=1.75$, \sgn{\mu}=+) and varied the messenger scale $M_{\text{mess}}$ and the parameter $\Lambda=\langle F_S \rangle / \langle S\rangle$. Here, $F_S$ and $S$ denote the auxiliary and scalar components of the gauge singlet field that is responsible for SUSY breaking.

Unfortunately, the full parameter space of minimal gauge mediation is ruled out for any of the fourth generation masses that are compatible with experiment. The small region where only the lightest $\widetilde{\tau}'$ is tachyonic (colored red in Fig.~\ref{fig:gGMSB}) can possibly be tractable in the sense that additional model building assumptions may lift the mass into the positive regime. This, however, would necessarily depart from the elegant and simple picture of minimal gauge mediation.

\section{Conclusions}

A fourth chiral generation is not favored by experiment, but it is not ruled out either. As an obvious extension of the Standard Model, it should be considered as a possibility for new physics to be discovered at the LHC. In the Standard Model, the presence of the fourth generation fermions with appropriately chosen masses may ease the tension between the lower experimental bound on the Higgs mass and its best-fit value from electroweak precision data. In the context of SUSY breaking, however, a fourth generation is problematic: Two of the most popular mechanisms for SUSY breaking do not work, and the main problem is the loss of perturbativity at scales.

\section*{Acknowledgments}

I thank the organizers of \textit{Les Rencontres de Moriond EW 2011} for providing an inspiring and hospitable atmosphere, Rohini M.~Godbole and Sudhir K.~Vempati for their collaboration, and Tom\'a\v{s} Je\v{z}o for comments.

\end{document}